\documentclass[preprint,aps,prc,superscriptaddress,showkeys]{revtex4-2}
\usepackage{graphicx} 
\usepackage{amsmath}
\usepackage{appendix}
\usepackage{physics} 
\usepackage{array} 
\usepackage{float}
\usepackage{soul}        
\usepackage{xcolor}
\sethlcolor{yellow}

\usepackage{fancyvrb}

\begin{document}
\title{E0 transition strengths as a tool to constraint model parameters. Application to even-even Xe isotopes}
\author{Pablo Mart\'in-Higueras}
\email{pablo.martin@dci.uhu.es}
\affiliation{Departamento de Ciencias Integradas y Centro de Estudios Avanzados en F\'isica, Matem\'atica y Computaci\'on, Universidad de Huelva, 21071 Huelva, Spain}

\author{Jos\'e-Enrique Garc\'ia-Ramos}
\email{enrique.ramos@dfaie.uhu.es}
\affiliation{Departamento de Ciencias Integradas y Centro de Estudios Avanzados en F\'isica, Matem\'atica y Computaci\'on, Universidad de Huelva, 21071 Huelva, Spain}
\affiliation{Instituto Carlos I de F\'{\i}sica Te\'orica y Computacional, Universidad de Granada, Fuentenueva s/n, 18071 Granada, Spain}

\begin{abstract}
\begin{description}
\item[Background]
In any nuclear structure model, the accurate description of E0 transition rates constitutes a significant challenge; at the same time, these observables provide a stringent constraint on the admissible parameter space of the model.

\item[Purpose]  
This work pretend to study the value of certain $\rho^2(E0)$ values or ratios over the whole parameter space of the interacting boson model. As an application, the case of the even-even Xe isotopes will be explored.

\item[Method]  
The interacting boson model will be considered for the calculation of $\rho^2(E0)$ values for some key transitions and to explore how these observables change over the parameter space of the model. 

\item[Results]  
Several key $\rho^2(E0)$ values or ratios are considered and contour plots in the parameter space of the interacting boson model (Casten triangle) are computed. The results for even-even Xe isotopes are superimposed on the contour plots. 

\item[Conclusions]  
The presented analysis confirms that $\rho^2(E0)$ values and, in particular, certain ratios allow to constraint the parameter space of the interacting boson model. There are certain regions where the $\rho^2(E0)$ values cannot be altered by the fine tuning of the Hamiltonian parameters.
\end{description}
\end{abstract}

\date{\today}
\maketitle

\section{Introduction}

Electric monopole (E0) transition have long been a topic of interest in nuclear physics because it provides a sensitive test for nuclear models, e.g., nuclear shape-coexistence, quantum phase transition or isomeric shift \cite{Wood99}. The E0 operator can only connect states of the same angular momentum and, as a matter of fact, it couples the nucleus to the atomic electrons, being predominantly mediated by internal conversion electrons \cite{Church1956}. Other possible process, if enough energy is available, is the creation of positron-electron pairs, while the emission of two photons is highly hindered. Of course, a single photon emission is forbidden due to angular momentum considerations.  

The transition probability involving a E0 operator is connected to the matrix element of the Coulomb interaction between the initial and the final states \cite{Church1956},
\begin{equation}
\label{hamilt_E0}
    \langle \psi_i |H_C| \psi_f \rangle  = \langle \psi_i |\sum_{p,e}\frac{-\alpha}{|r_p-r_e|}| \psi_f \rangle 
\end{equation}

The transition probability, $W(E0)$, associated to E0 transitions can be factorized into two pieces, one depending only on nuclear properties while the other on electronic ones,
\begin{equation}
\label{W_E0}
    W(E0)=\frac{1}{\tau(E0)}=\rho^2(E0)\times \Omega_{IC},
\end{equation}
where $\tau(E0)$ is the mean-life with respect to E0 transitions, $\Omega_{IC}$ is a factor depending on the different electronic shells of the considered nucleus and on the transition energy,
\begin{equation}
    \Omega_{IC}=\Omega_K(E0)+\Omega_{L_1}(E0)+\ldots+\Omega_\pi,
\end{equation}
where the very last term corresponds to the pair creation contribution. Finally, $\rho^2(E0)$ is dimensionless and depends on nuclear properties, 
\begin{equation}
\label{rhoE0}
\rho^2(E0)=\left |\frac{1}{e R^2}\sum_k\bra{\psi_f} e_k r_p^2\ket{\psi_i}\right |^2, 
\end{equation}
where $R=1.2 A^{1/3}$ fm, $e_k$ are effective charges and the sum goes over all nucleons.

Once defined Eq.~(\ref{rhoE0}), it is natural to introduce the E0 transition operator as,
\begin{equation}
\label{TE0}
\hat{T}(E0)=\sum_k e_k r_k^2.
\end{equation}
Moreover, it is also possible to define a charge radius operator as,
\begin{equation}
\label{T_r2}
    \hat{T}(r^2)=\frac{\sum_k e_k r_k^2}{Z e_p +N e_n} = \frac{\hat{T}(E0)}{Z e_p +N e_n}, 
\end{equation}
where $e_p$ and $e_n$ are the effective charges for protons and neutrons. The basic assumption is that the effective charges for the $\hat{T}(E0)$ and the $\hat{T}(r^2)$ operators are the same \cite{Zerg12}. Therefore,
\begin{equation}
 \label{rhoE0_bis}
\rho^2(E0)= \left |\frac{\bra{\psi_f} \hat{T}(E0)\ket{\psi_i}}{e R^2}\right |^2 =
\left | \frac{(Z e_p +N e_n)\bra{\psi_f}\hat{T}(r^2)\ket{\psi_i}}{e R^2}\right |^2.
\end{equation}

The experimental determination of $\rho^2(E0)$ is a really challenging task and the available experimental information is really scarce as shown in the compilations \cite{Wood99,Kibe05}. However, as already said, its knowledge can shed light on the particular structure of the nuclear wave function and on its evolution along a chain of isotopes. The transitional regions where the nuclear shape chain rapidly or the coexistence of several configurations with different shape plays a major role are ideal areas where the knowledge of E0 transition rates can help to refine nuclear models.

To explore the capability of E0 transition rates as a key observable to constraint nuclear models, we will consider the even-even Xe isotopes. In recent years, the study of Xe isotopes has become of interest. In particular, because $^{124-126}$Xe \cite{Barr2014} and $^{134-136}$Xe \cite{Barr2014,Joki2023} are double-$\beta$ decay nuclei. Moreover, Xe isotopes are nuclei close to $\gamma$-unstable shapes and that, therefore, preserve the O(6) dynamical symmetry of the interacting boson model (IBM) \cite{iach87}. However,  measurements carried out for $^{124-126-128}$Xe suggest that the O(6) symmetry is badly broken, but the O(5) is preserved \cite{Wern2001,Rain2010}. Hence, it is of interest to explore the value of $\rho^2(E0)$ in even-even Xe isotopes to try to unravel its O(6) character. 

This work is organized as follows. In Section \ref{sec-IBM}, we present the main features of the IBM and we explain how to compute E0 transition rates and radii. In Section \ref{sec-Xe-fit}, we show an IBM calculation for $^{114-134}$Xe using the more up-to-date experimental information to fix the Hamiltonian parameters. In Section \ref{sec-E0-calc}, relevant $\rho^2(E0)$ values and ratios are studied in the parameter space of the IBM and applied to the case of even-even Xe isotopes. Finally, in Section \ref{sec-conclu}, our summary and conclusions are depicted. 

\section{The interacting boson model and the E0 operator}
\label{sec-IBM}
The IBM is one of the most extensively studied algebraic model for describing nuclear properties \cite{iach87}. Its central assumption is that protons and neutrons couple in pairs to form bosons with angular momentum $L=0$ and $L=2$, known as $s$ and $d$ bosons, respectively. These bosons behave collectively, similarly to Cooper pairs, and their existence is the consequence of the strong pairing interaction in nuclei between alike particles. The most appealing feature of the model is its algebraic structure and, as a matter of fact, it has U(6) as a dynamical algebra. Hence, the Hamiltonian and any other operator of the model will be written in terms of the generators of the U(6) Lie group.

The Hamiltonian of the IBM can be written in several forms, but we will focus on the multipolar one and, in particular, on a reduced version of it, called the extended consistent-Q formalism (ECQF) \cite{warner83}, which Hamiltonian operator is written as
\begin{equation}
    H = \epsilon_d\ \hat{n}_d + \kappa' \hat{L}\cdot \hat{L}+ \kappa\ \hat{Q}(\chi)\cdot \hat{Q}(\chi), 
    \label{H_general}
\end{equation}
where $\cdot$ stands for the scalar product,  $\hat{n}_d$ is the d boson number, $\hat{L}$ the angular momentum, and $\hat{Q}(\chi)$ the quadrupole operator,
\begin{align}
  & \hat{n}_d=(d^\dagger \cdot \widetilde{d}), \\
  & \hat{L}=\sqrt{10}[d^\dagger\times \widetilde{d}]^{(1)}, \\
  & \hat{Q}(\chi) =  [d^\dagger \times \widetilde{s}+s^\dagger\times \widetilde{d}]^{(2)}+\chi[d^\dagger\times \widetilde{d}]^{(2)}.
\end{align}
An extra element of the ECQF  is that the E2 transition operator has the same structure as the $\hat{Q}(\chi)$ operator appearing in the Hamiltonian,
\begin{equation}
\label{TE2}
    \hat{T}(E2)= e_{eff}\ \hat{Q}(\chi).
\end{equation}
Therefore, within the ECQF only 5 parameters are needed to describe the excitation energies and E2 transition rates of the positive parity states of a given nucleus.

The parameters of the Hamiltonian and E2 operators are somehow elements external to the model. They can be mapped from the Shell Model \cite{Schol1985} or from energy density functionals \cite{nomura08}. An alternative is to use a phenomenological approach in which the parameters are fixed through a least-squares fit for reproducing as best as possible excitation energies or electromagnetic transition rates among other observables \cite{Garc09}. In this work we will use the latter approach.   

The IBM allows the description of the nuclear charge radius,  
\begin{equation}
    \hat{T}(r^2)= \expval{r^2}_0 + \gamma \hat{A}+\beta\hat{n}_d,
\label{T_r2_ibm}
\end{equation}
where $\hat{A}$ is the mass number operator (equivalently, the number of boson operator, $\hat{N}_b$, can be also used) and $\expval{r^2}_0$ is just the charge radius of the core nucleus. Therefore, the radius will present a global linear trend but shaped by the local deformation introduced by the $\hat{n}_d$ operator.

The E0 operator according to Eq.~(\ref{T_r2}) is written as, 
\begin{equation}
    \hat{T}(E0)= (Z e_p +N e_n) \beta\hat{n}_d,
    \label{trans_e0_eq}
\end{equation}
where the diagonal contributions, $\expval{r^2}_0$ and $\gamma \hat{A}$, are excluded because they vanish for transitions involving different states. Once more, we consider the basic assumption that the coefficients of both operators are the same.

\section{The description of $^{114-134}$Xe isotope chain with the IBM}
\label{sec-Xe-fit}
\subsection{Excitation energies and $B(E2)$ reduced transition probabilities}
In this section, we will conduct an analysis of the excitation energies and E2 transition rates of $^{114-134}$Xe to fix the parameters appearing in the Hamiltonian (\ref{H_general}) and the effective charge of the $\hat{T}(E2)$ operator (\ref{TE2}). In \cite{Puddu1980}, this region was already studied using the IBM-2 \cite{arima77}, while more recently, in \cite{Baid2024} a Bohr Hamiltonian with a $\gamma$-unstable and $\beta^6$ potential was considered. 

We will use as core for protons Z=50 and that neutrons are in shell N=50-82. Therefore, we will describe 11 nuclei around mid-shell for neutrons.
\begin{figure}[hbt]
\includegraphics[width=0.9\textwidth]{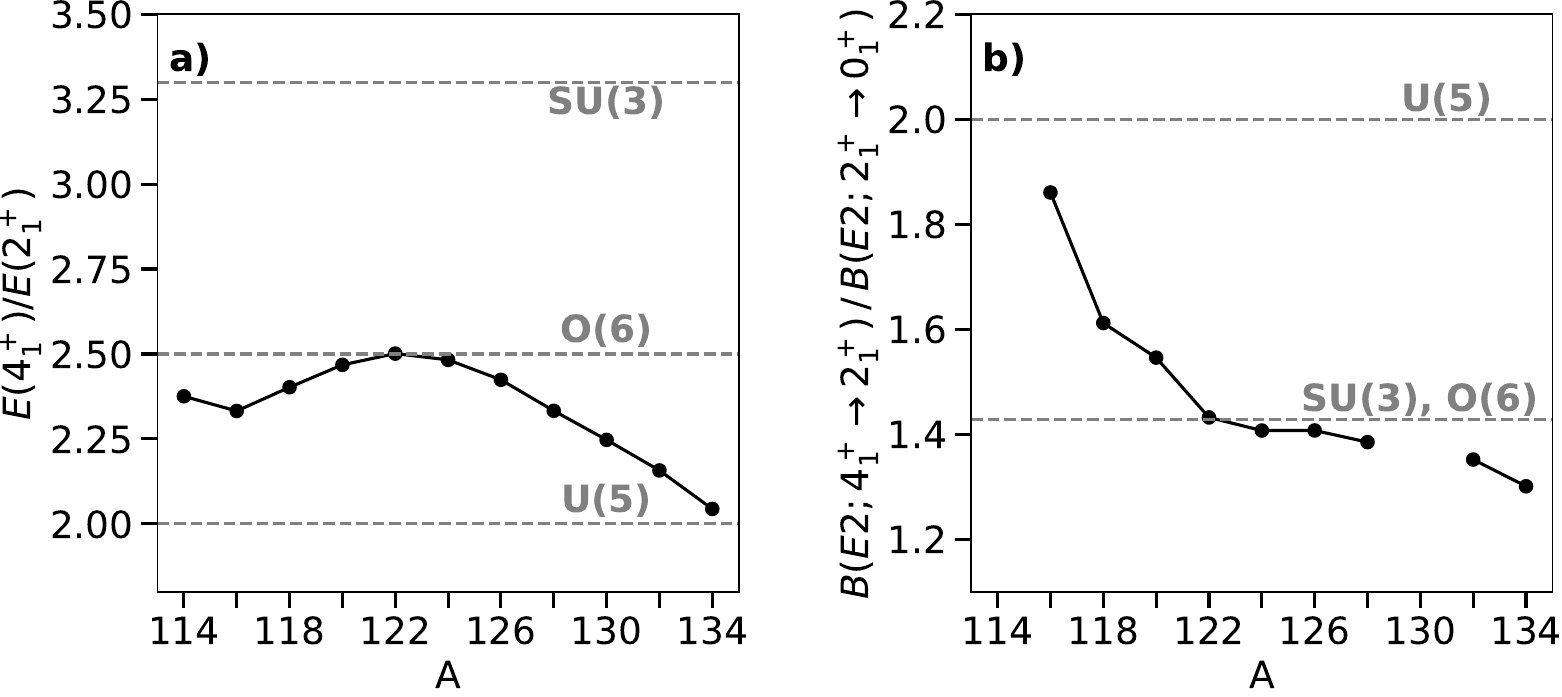}
\caption{a) Energy ratio $E(4_1^+)/E(2_1^+)$ and b) $B(E2: 4_1^+\rightarrow 2_1^+)/B(E2: 2_1^+\rightarrow 0_1^+)$ as a function of A. The dynamical symmetry limits  are given as a reference.} 
\label{fig-ratios_exp}
\end{figure}

Before starting with the IBM description of the isotope chain, it is worthy to have a detailed view of the energy and B(E2) ratios involving the lowest states of the yrast band, becasue they have a well defined value at the dynamical symmetry limits of the IBM. The corresponding experimental values are  presented for the whole even-even Xe isotope chain in Fig.~\ref{fig-ratios_exp}, together with the U(5), SU(3), and O(6) values. This figure confirms that Xe isotopes are close to the $\gamma$-soft case and clearly they are far from a $\gamma$-rigid situation (SU(3)). This is specially evident around mid-shell, $^{120}$Xe. 
Note that, specially at the beginning and at the end of the shell, the energy and the B(E2) ratios do not present a consistent behavior, one pointing to the U(5) limit, while the other to the O(6) one.
\begin{figure}[hbt]
    \centering
    \includegraphics[width=0.75\textwidth]{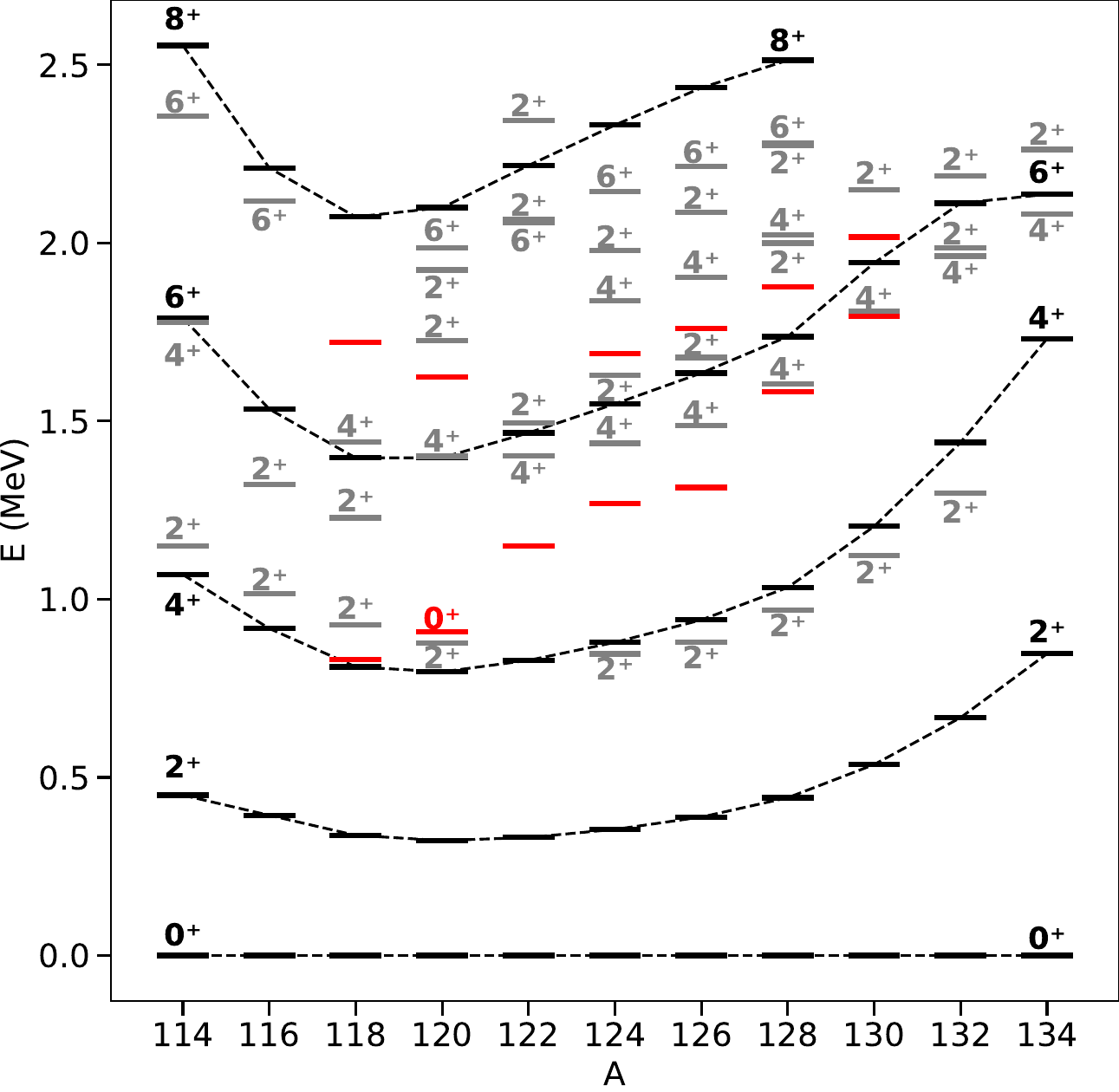}
    \caption{Experimental excitation energy systematics. Black lines for the yrast band, red lines for excited $0^+$ states and gray lines for the rest of levels.}
    \label{fig-sistematica_exp}
\end{figure}

The complete spectrum of the Xe isotope chain up to $2.5$ MeV is shown in Fig.~\ref{fig-sistematica_exp}. The yrast band is displayed together with several representative levels from the yrare bands. The most striking feature is the overall compression of the level scheme as one moves toward mid-shell. Concerning the $0^+_2$ state, one sees that it initially appears as part of the $4_1^+-2_2^+-0^+_2$ ($n_d=1$) vibrational triplet at mid-shell, and then shifts to higher energy, eventually forming the $6^+_1-4_2^+-0^+_2$ ($\sigma=\tau=3$) $\gamma$-soft triplet near the end of the shell.


One can think in the $0_2^+$ state as a member of a intruder band, however the observed parabolic systematics is rather flat and there are not other clear evidences regarding the presence of intruder states in Xe isotopes \cite{heyde11,Garr22}. Therefore, IBM with a single configuration seems to be a good approximation to describe even-even Xe isotopes.

The procedure to describe Xe isotopes is to carry out a fitting procedure, trying to obtain the best possible overall agreement with the experimental data including
both the excitation energies and the B(E2) reduced transition probabilities. We will use Eqs.~(\ref{H_general}) and (\ref{TE2}) and, therefore, 5 parameters are needed for each nucleus. We impose as an extra constraint to obtain parameters that change smoothly passing from isotope to isotope. We closely follow Ref.~\cite{Garc14b} using evaluated experimental data taken from \cite{ENSDF}.
\begin{figure}[hbt]
  \centering
  \includegraphics[width=0.6\linewidth]{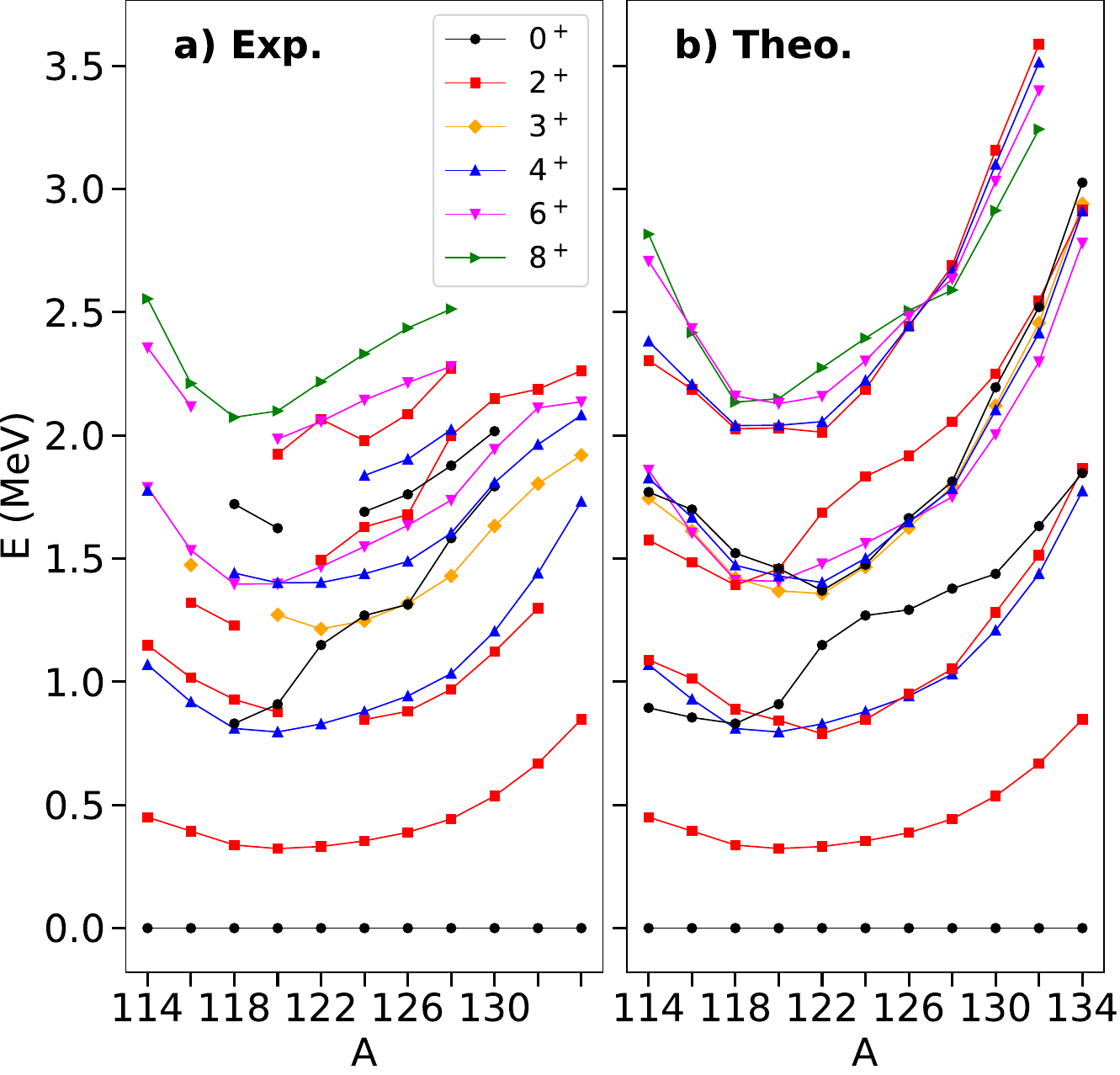}
  \caption{Experimental excitation energies (panel a)) and the theoretical results (panel b)),  obtained from the IBM.} 
  \label{fig-energ-comp}
\end{figure}

\begin{table}[hbt]
\caption{Hamiltonian and $\hat{T}(E2)$ parameters resulting from the present study.  All quantities have the dimension of energy (given in keV), except $\chi$, which is dimensionless and $e_{eff}$ which is given in units $\sqrt{\mbox{W.u.}}$ The used number of bosons is also provided.}
\label{tab-fit-parameters}
\begin{center}
\begin{ruledtabular}
\begin{tabular}{cccccc}
Nucleus (N$_B$)& $\varepsilon$& $\kappa$& $\chi$& $\kappa'$& $e_{eff}$\\
\hline
 $^{114}$Xe (7)&   602.5&   -19.99& -0.97&    9.75&   2.31\footnotemark[1]\\ 
 $^{116}$Xe (8)&   602.3&   -19.28& -0.89&    5.95&   2.31\\
 $^{118}$Xe (9)&   567.3&   -19.22& -0.65&    4.38&   2.22\\
 $^{120}$Xe (10)&  582.5&   -21.17& -0.44&    4.66&   2.20\\ 
 $^{122}$Xe (9)&   532.1&   -29.41& -0.15&    4.70&   1.99\\
 $^{124}$Xe (8)&   523.4&   -35.03& -0.13&    3.64&   1.86\\
 $^{126}$Xe (7)&   543.5&   -38.04& -0.20&    1.80&   1.82\\ 
 $^{128}$Xe (6)&   572.5&   -40.00& -0.00&   -1.50&   2.18\\
 $^{130}$Xe (5)&   650.0&   -40.00& -0.31&   -3.00&   2.27\\
 $^{132}$Xe (4)&   731.0&   -36.88& -0.11&  -5.19&   2.13\\ 
 $^{134}$Xe (3)&   881.9&   -30.00& -0.78&   -4.00&   2.13\\ 
\end{tabular}
\end{ruledtabular}
\end{center}
\footnotetext[1]{Effective charge taken from $^{116}$Xe.}
\end{table}



\begin{figure}[h]
  \centering
  \includegraphics[width=0.6\linewidth]{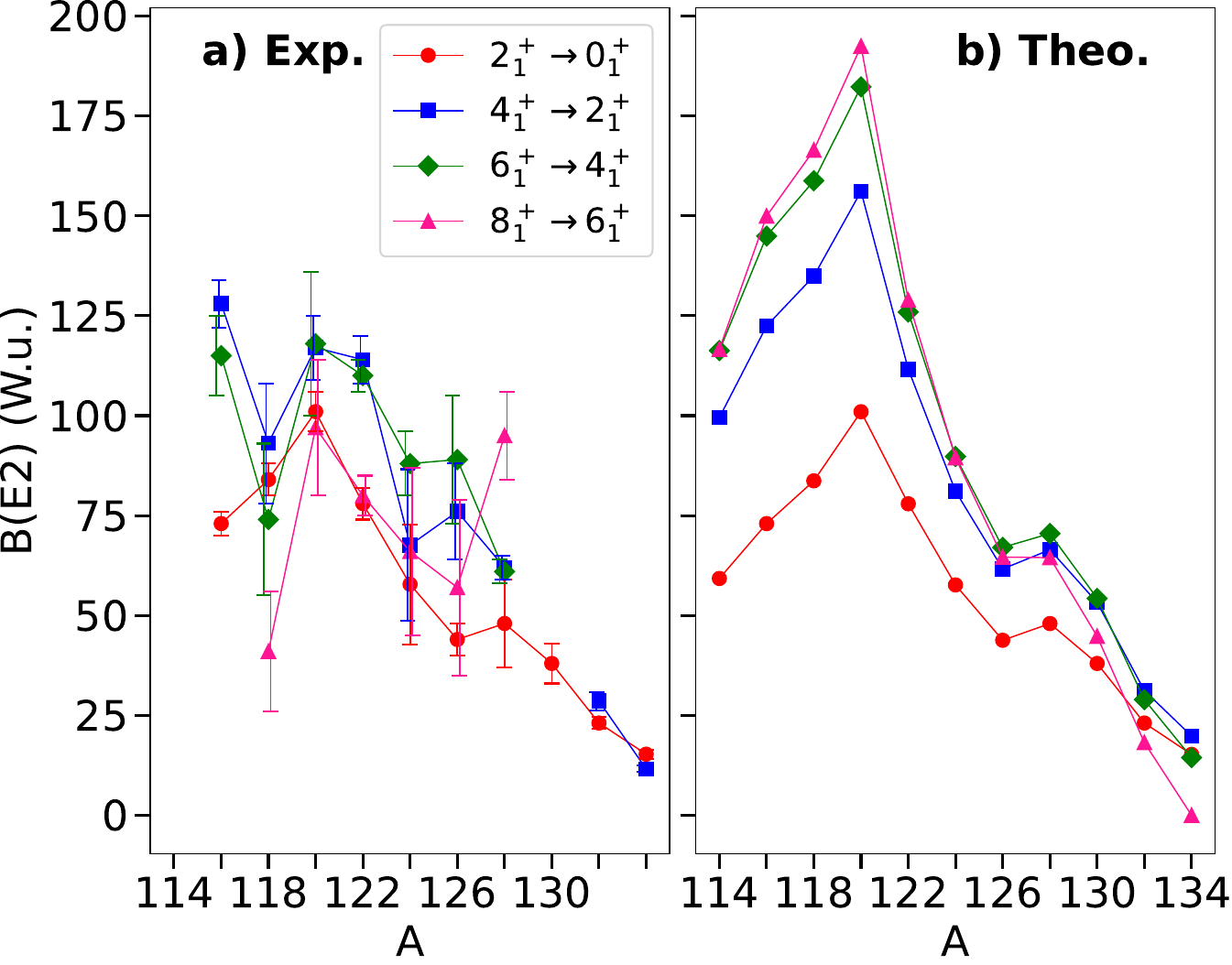}
  \caption{Comparison of the absolute $B(E2)$ reduced transition probabilities along the yrast band, given in W.u. Panel a) corresponds to experimental data and panel b) to the theoretical IBM results.} 
  \label{fig-be2-1}
\end{figure}
\begin{figure}[hbt]
  \centering
  \includegraphics[width=0.6\linewidth]{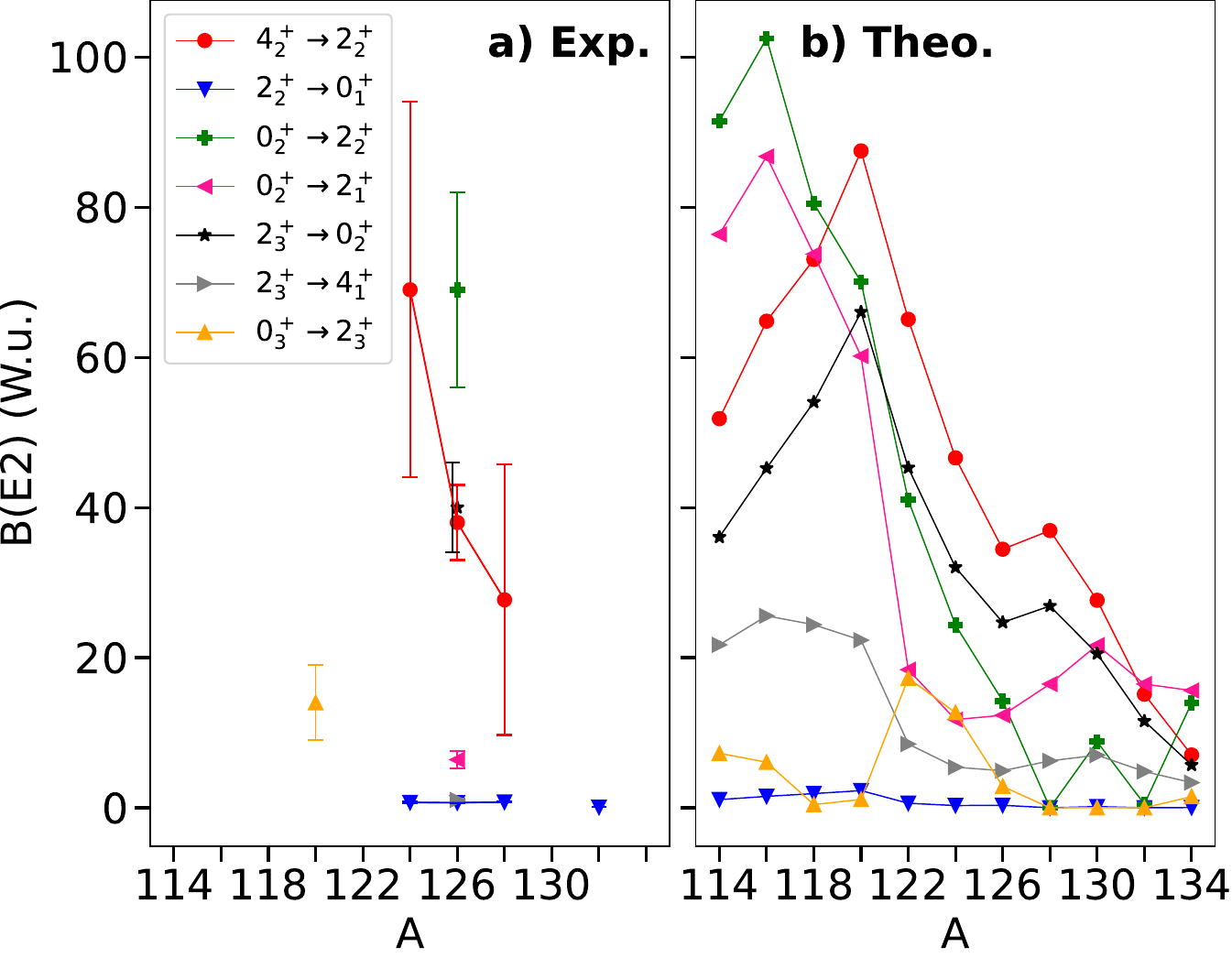}
  \caption{Same as Fig.~\ref{fig-be2-1} but for few intraband transitions.}
  \label{fig-be2-2}
\end{figure}

The obtained values of the Hamiltonian and E2 parameters are given in Table \ref{tab-fit-parameters}, while the comparison between experimental data and IBM results is provided in Figs.~\ref{fig-energ-comp}, \ref{fig-be2-1} and \ref{fig-be2-2} for excitation energies and E2 transition rates, respectively. From Table \ref{tab-fit-parameters}, one can conclude that Xe isotopes does not own O(6) dynamical symmetry, because they have a relatively large contribution from the $\hat{n}_d$ term. However, in the second part of the shell, the nuclei present a behavior rather close to the O(5) symmetry with values of $\chi$ close to zero. 

Regarding the excitation energies, Fig.~\ref{fig-energ-comp} compares the theoretical predictions with the experimental data. Overall, the agreement is reasonable, although several deficiencies are apparent. The first excited $0^+$ states are correctly reproduced up to mid-shell. However, near the end of the shell, the theoretical $0_2^+$ level forms a triplet with the $2_2^+$ and $4_1^+$ states, whereas experimentally the corresponding triplet involves the $6_1^+$ and $4_2^+$ states. More generally, toward the end of the shell, the theoretical spectrum exhibits a fully vibrational character, while the experimental spectrum shows significantly stronger mixing between multiplets and a more compressed level spacing.

The theoretical and experimental intraband E2 transition rates are shown in Fig.~\ref{fig-be2-1} for comparison. The value of the $2_1^+\rightarrow 0_1^+$ transition is well reproduced, exhibiting a maximum of collectivity in the middle of the shell and a minimum toward its end. It should be emphasized that, in the fitting procedure, this theoretical transition was constrained to follow the experimental value. While the overall trend of the remaining transitions is also correctly captured, the theoretical maxima are significantly larger than the experimental ones.

Regarding the intraband transitions, theoretical and experimental values are shown in Fig.~\ref{fig-be2-2}. It should be emphasized that the available experimental data are very limited, which makes it difficult to reliably assess the quality of the fit. Nonetheless, a qualitative level of agreement is achieved.

\subsection{Nuclear charge radii and $\rho^2(E0)$ values}
Once the Hamiltonian parameters are determined, it is very convenient to explore the behavior of nuclear observables that involve the Xe nuclear chain as a whole, such as the nuclear charge radius. To this end, we will consider the Hamiltonian parameters previously  determined. This will also serve to check the reliability of the obtained parameters.

\begin{figure}[hbt]
    \centering
    \includegraphics[width=0.8\textwidth]{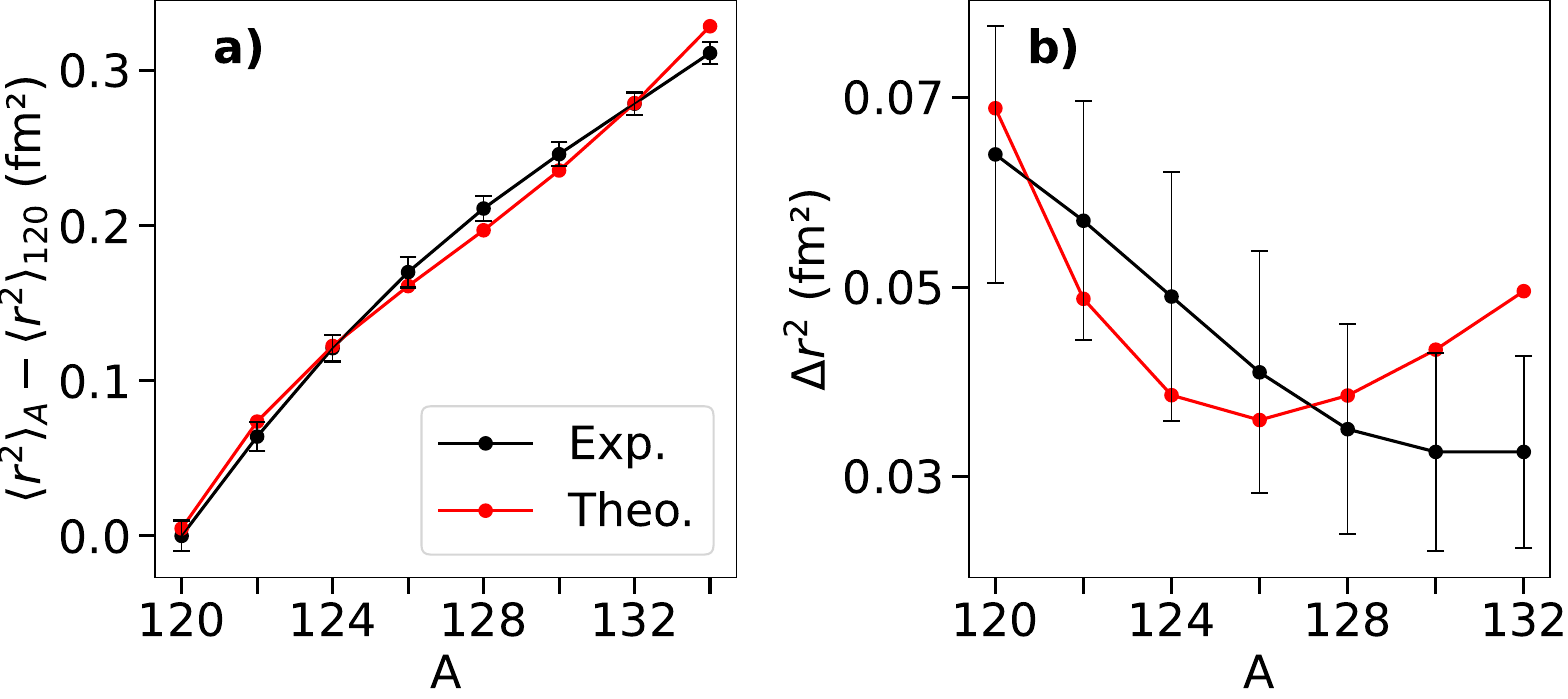}
    \caption{Comparison of experimental and IBM result for a) charge radii and b) isotopic shift.}
    \label{fig-radii}
\end{figure}

We will employ Eq.~(\ref{T_r2_ibm}), so that the radius is expressed as
\begin{equation}
    \expval{r^2}= \expval{r^2}_0 + \gamma A+\beta\expval{\hat{n}_d},
\label{radius}
\end{equation}
where the coefficients are kept constant for the whole isotope chain. We will only focus on the second part of the shell. 
\begin{figure}[hbt]
    \centering
    \includegraphics[width=0.8\linewidth]{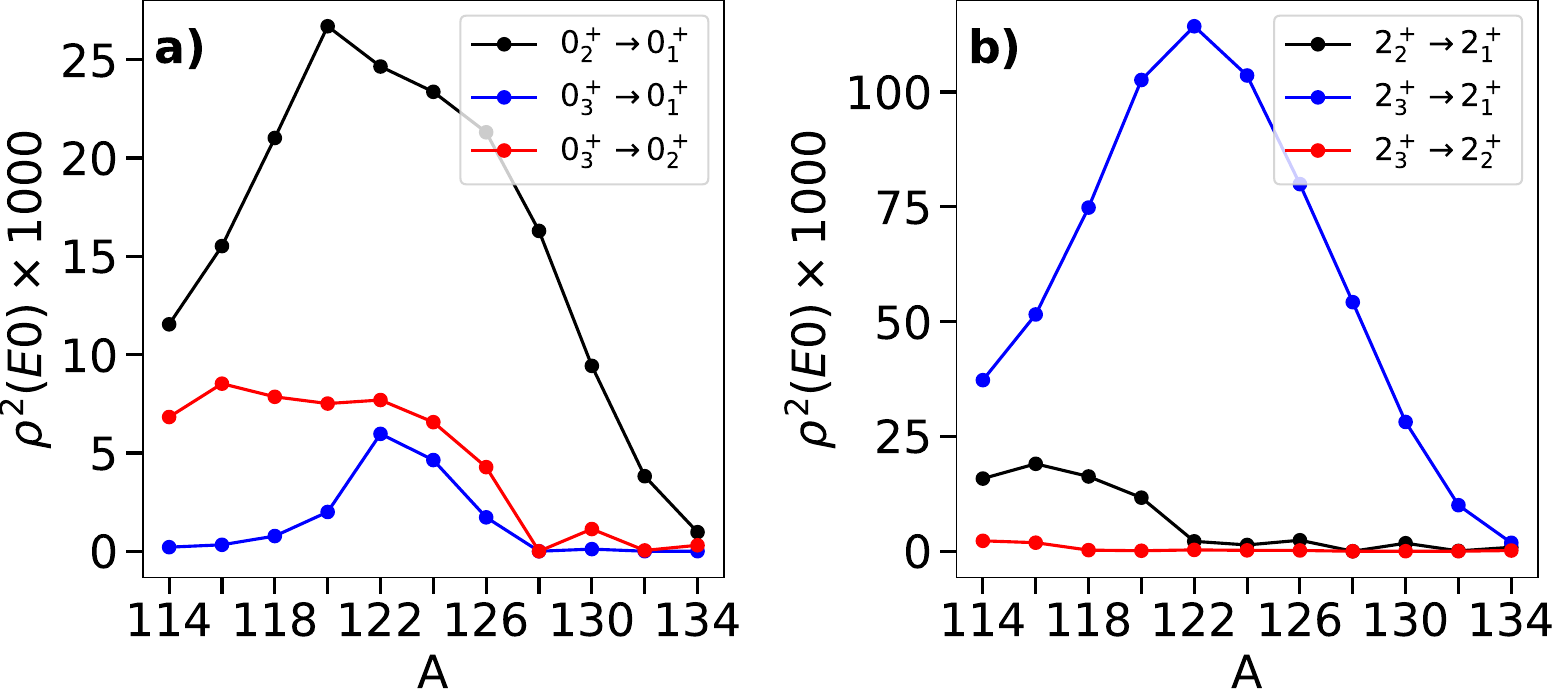}
    \caption{Theoretical values of $\rho^2(E0)$ for Xe isotopes. Panel a) for $0^+$ and b) for $2^+$ states.}
    \label{rho2result}
\end{figure}

The obtained parameters are $\expval{r^2}_0=(-3.3\pm0.4)\ \text{fm}^2$, $\gamma=(0.027\pm 0.003)\ \text{fm}^2$ and $\beta=(0.05\pm 0.03)\ \text{fm}^2$. In Fig.~\ref{fig-radii} theoretical and experimental results (taken from \cite{Ange13}) are compared, considering as a reference the radius of $^{120}$Xe. In panel a) the radius is presented, while in panel b) the isotopic shift, $\Delta \expval{r^2}=\expval{r^2}_{A+2}-\expval{r^2}_{A}$, is shown. One can see a rather smooth trend without abrupt changes over the whole isotope chain. The agreement is good, in general, although some discrepancies are observed at the end of the shell, which are enhanced when presenting the isotopic shift. 

As soon as the parameters of the radius operator  are determined using Eqs.~(\ref{T_r2_ibm}) and (\ref{radius}), it is possible to compute $\rho^2(E0)$ values using Eq.~(\ref{rhoE0_bis}). Assuming the standard values $e_p= 1 e$ and $e_n=0.5 e$ one can get the systematics for the transitions involving the lower $0^+$ and $2^+$ states, as shown in Fig.~\ref{rho2result}. Note that in this case only the value of $\beta$ is relevant. The main features of this figure are that the maxima appear around the mid-shell, that $\rho^2(0_2^+\rightarrow 0_1^+) \gg \rho^2(0_3^+\rightarrow 0_1^+)$ while $\rho^2(2_3^+\rightarrow 2_1^+) \gg \rho^2(2_2^+\rightarrow 2_1^+)$, and that $\rho^2(E0)$ for $2^+$  are $4-5$ times larger than for $0^+$ states. As we will see in next section, these features strongly depend on the Hamiltonian of the isotope chain.

\section{Dependence of $\rho^2(E0)$ on the IBM parameter space}
\label{sec-E0-calc}
Up to this point, we have reported the $\rho^2(E0)$ values for the Xe isotopes. However, the range of possible $\rho^2(E0)$ values permitted by different choices of the IBM Hamiltonian parameters remains unclear. In particular, it is not evident whether the values obtained for the Xe isotope chain reflect a generic behavior of the model or if they are instead highly specific to this set of nuclei.

\begin{figure}[hbt]
    \centering
    \includegraphics[width=0.9\linewidth]{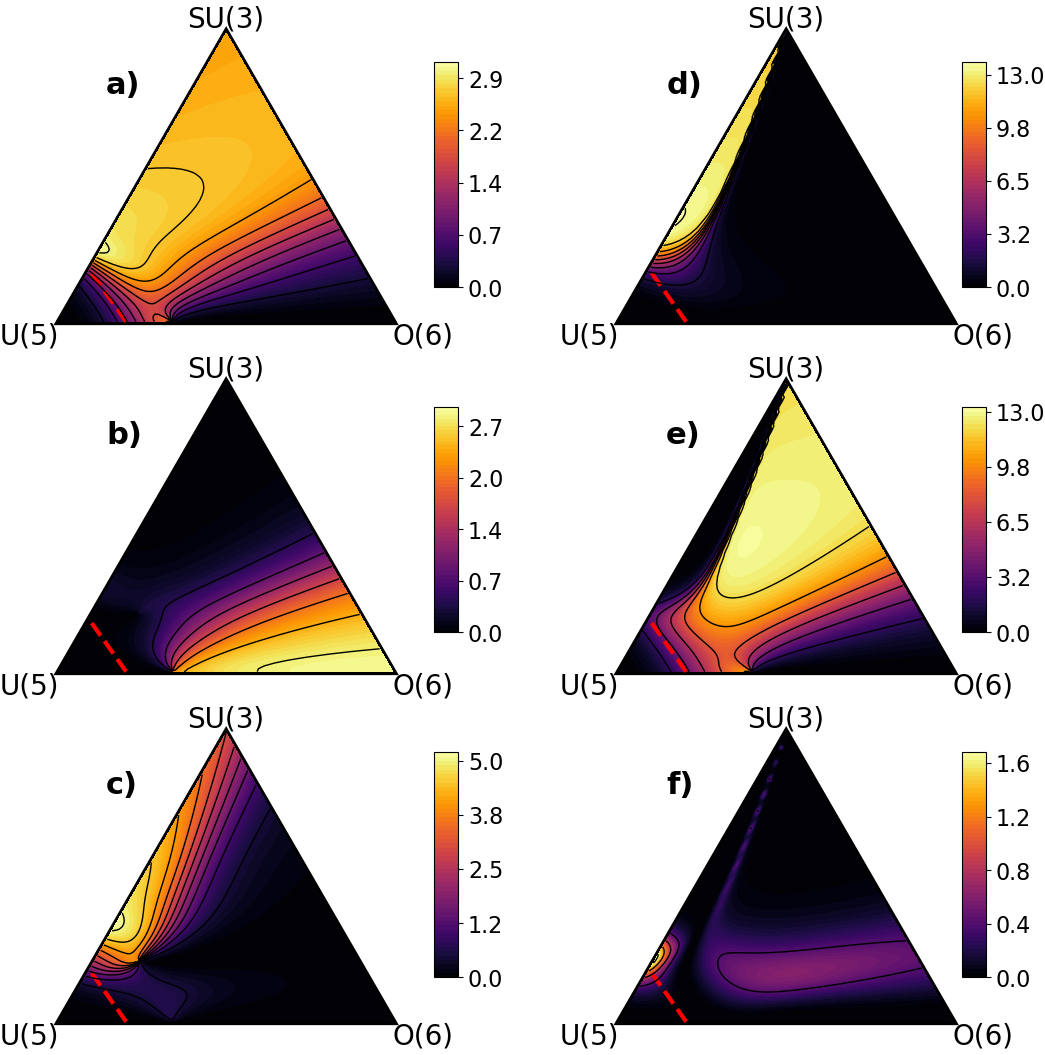}
    \caption{$|\bra{\psi_f} \hat{n}_d \ket{\psi_i}|^2$ values for selected states as a function of the ECQF Hamiltonian parameters for $N_B=10$. a) $0_2^+\rightarrow 0_1^+$, b) $0_3^+\rightarrow 0_1^+$, c) $0_3^+\rightarrow 0_2^+$, d) $2_2^+\rightarrow 2_1^+$, e) $2_3^+\rightarrow2_1^+$, and f) $2_3^+\rightarrow2_2^+$. The symmetry limits and the QPT line are given for reference.}
    \label{fig-surfaces_E0}
\end{figure}

\begin{figure}[hbt]
\centering
\begin{minipage}{0.8\linewidth}
    \centering
    \includegraphics[width=\linewidth]{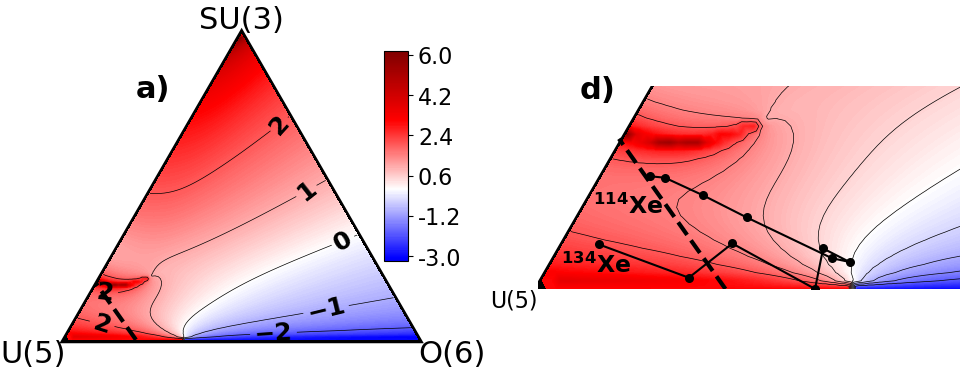}
    \vspace{0.1cm}
    \includegraphics[width=\linewidth]{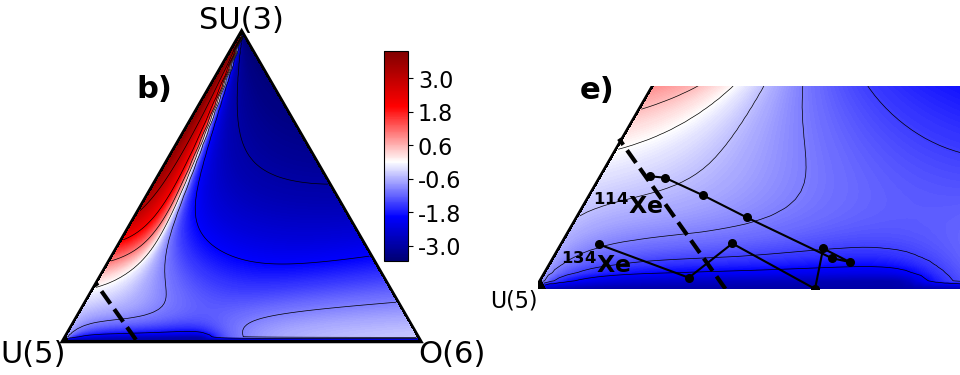}
    \vspace{0.1cm}
    \includegraphics[width=\linewidth]{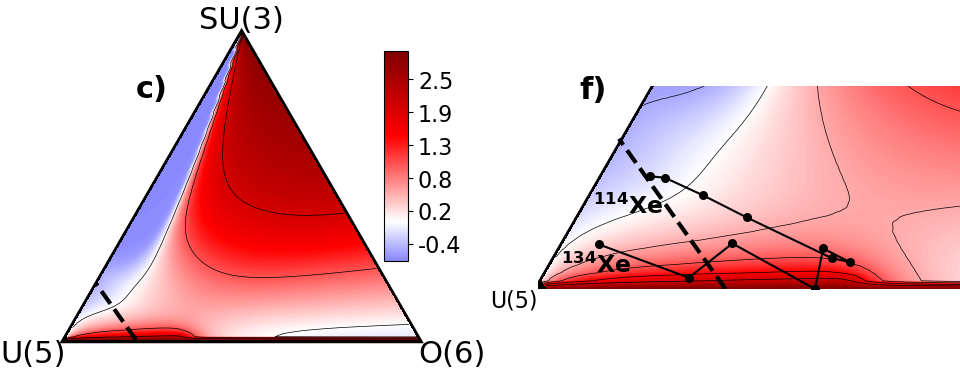}
\end{minipage}
\caption{a) $\log\left(\frac{\rho^2(E0:\ 0_2^+\rightarrow 0_1^+)}{\rho^2(E0:\ 0_3^+\rightarrow 0_1^+)}\right)$, 
b)  $\log\left(\frac{\rho^2(E0:\ 2_2^+\rightarrow 2_1^+)}{\rho^2(E0:\ 2_3^+\rightarrow 2_1^+)}\right)$, c)  $\log\left(\frac{\rho^2(E0:\ 0_2^+\rightarrow 0_1^+)}{\rho^2(E0:\ 2_2^+\rightarrow 2_1^+)}\right)$, d), e) and f) stand for the corresponding zoomed figures, including the position of Xe isotopes. The symmetry limits and the QPT line are given for reference.}
\label{fig-ratioCasten_isotopes}
\end{figure}

In order to explore the whole parameter space of the IBM Hamiltonian, we will consider the ECQF Hamiltonian (\ref{H_general}) but excluding the $L^2$ dependence, because this term does not affect the wavefunction,  and rescaling the Hamiltonian to enhance the presence of quantum phase transitions (QPTs) \cite{Cejn09} in the so-called Casten triangle \cite{warner83}. Hence, we will write the Hamiltonian up to a scale factor as,   
\begin{equation}
    H = \xi\ \hat{n}_d - \frac{1-\xi}{N_B}\ \hat{Q}(\chi)\cdot \hat{Q}(\chi),
\label{H_squem}
\end{equation}
where $N_B$ stands for the number of bosons. This Hamiltonian has the advantage of presenting a QPT around $\xi=0.8$ in the large $N_B$ limit. In Figs.~\ref{fig-surfaces_E0} and \ref{fig-ratioCasten_isotopes}, we use the Casten triangle to present our results. In this representation, the symmetry limits correspond to the following parameters: U(5) ($\xi=1$), SU(3) ($\xi=0$ and $\chi=-\sqrt{7}/2$) and O(6) ($\xi=0$ and $\chi=0$).

First of all, we will explore the variation of $\rho^2(E0)$ on the parameter space studying $|\bra{\psi_f} \hat{n}_d \ket{\psi_i}|^2$ to avoid the dependence on the scale coefficient of the E0 operator. In particular, we will present the value of $|\bra{\psi_f} \hat{n}_d \ket{\psi_i}|^2$, on the one hand, for $0_1^+$, $0_2^+$, and $0_3^+$ and, on the other for $2_1^+$, $2_2^+$, and $2_3^+$ states.  In Fig.~\ref{fig-surfaces_E0}, we depict the corresponding values for $N_B=10$, assuming that the results do not strongly depend on $N_B$, except for very low $N_B$ values. The first evident fact is that it is possible to get non-vanishing values of $\rho^2(E0)$ without resorting to the mixing of several configurations as pointed out in \cite{Wood99}. This has been noted in several previous publications \cite{VonBrentano2004}, but generally only $0^+$ states were considered. Here, the same fact is observed for $2^+$ states. Anyhow, there are large areas in the parameter space where the corresponding $\rho^2(E0)$ values vanish. 

$\rho^2(0_2^+\rightarrow 0_1^+)$ is zero around the U(5) symmetry limit and in the area around the O(6) symmetry limit; $\rho^2(0_3^+\rightarrow 0_1^+)$ is only different from zero next to the O(6) symmetry limit; $\rho^2(0_3^+\rightarrow 0_2^+)$ only presents non-zero values in the U(5)-SU(3) leg, with the exception of the area next to the U(5) symmetry limit where it vanishes. 

Concerning the $2^+$ states,  $\rho^2(2_2^+\rightarrow 2_1^+)$ is only different from zero in the U(5)-SU(3) leg, with the exception of the area next to the U(5) symmetry limit where it vanishes; $\rho^2(2_3^+\rightarrow 2_1^+)$ is only different from zero in the areas far from the symmetry limits; $\rho^2(2_3^+\rightarrow 2_2^+)$ presents a null or very small value almost in the whole parameter space. Note that the values at the SU(3) symmetry limit are not well defined due to the degeneracy between the $2_2^+$ and $2_3^+$ states \cite{Piet1985}. It is noticeable the complementary character of panels d) and e) with a perfect coincidence of the regions around the U(5)-SU(3) leg, which suggests the interchange of order between the $2_2^+$ and $2_3^+$ states. As a matter of fact, the line where these two states get degenerated seems to be connected with the so called Alhassid-Whelan arc of regularity \cite{Alha1991,Alha1991b}, as suggested in \cite{Dennis2010}. Clearly, there is a crossing of two $2^+$ states which is a hint for the presence of an underlying symmetry \cite{Aria03}. The behavior of $\rho^2(E0)$ for the $0^+$ states points to the interchange of the position of two states that never cross and present level repulsion \cite{Aria03}. Finally, it is evident, that the obtained values are much larger for $2^+$ than for $0^+$ states, as was noted previously.

In general, in regions where $|\bra{\psi_f} \hat{n}_d \ket{\psi_i}|^2$ varies only weakly, the matrix element becomes largely insensitive to the Hamiltonian parameters. Conversely, in regions where it changes rapidly, the dependence on the Hamiltonian parameters is strong. Consequently, in domains of slow variation, experimental information on this quantity provides little leverage to constrain the Hamiltonian: if the measured value is not reproduced, the model offers essentially no parametric flexibility to remedy the discrepancy. By contrast, in regions of steep variation, small adjustments of the Hamiltonian parameters can induce substantial changes in the matrix element. In such cases, the model can accommodate a broad range of experimental values with minimal impact on the overall energy spectrum.


In order to obtain $\rho^2(E0)$ values independent on the effective charges, it is convenient to consider the ratio of $\rho^2(E0)$ values. Moreover, this quantity allows to enhance the already noticed facts observed in Fig.~\ref{fig-surfaces_E0}. 

In Fig.~\ref{fig-ratioCasten_isotopes}, $\rho^2(E0)$ ratios are presented. In panel a) $\log(\frac{\rho^2(E0:\ 0_2^+\rightarrow 0_1^+)}{\rho^2(E0:\ 0_3^+\rightarrow 0_1^+)})$ over the whole Casten triangle is depicted, while in panel d) the position of the Xe isotopes are superimposed to the same results but in a more reduced area around the critical line. The $\rho^2(E0)$ ratio moves over 9 order of magnitude, but present a relatively smooth behavior. There is a narrow region surrounding the O(6) symmetry limit where the ratio is equal to one. One can see that Xe isotopes are relatively close to the QPT critical line, having $^{114}$Xe a vibrational character, evolving into a O(6) one at the mid-shell and returning to the vibrational one in $^{134}$Xe. The observed variation for the Xe isotopes is relatively modest. In this figure, we can also see that isotopes near the middle of the shell are close to a ratio of 1. Hence, a small change in the Hamiltonian parameters can shift the ratio to values slightly below or above 1. As a consequence, access to experimental information around mid-shell can significantly aid in constraining the parameters of the Hamiltonian.

In panels b) and e) $\log\left(\frac{\rho^2(E0:\ 2_2^+\rightarrow 2_1^+)}{\rho^2(E0:\ 2_3^+\rightarrow 2_1^+)}\right)$ is depicted. Once more, in panel e) the position of Xe isotopes is superimposed. In this case, the ratio moves over 6 orders of magnitude. The area between the U(5)-SU(3) leg and the arc of regularity shows a very rapid variation (4 orders of magnitude) in a relatively small space. This behavior implies that a wide range of experimental values exceeding unity, i.e., $\rho^2(E0:\ 2_2^+\rightarrow 2_1^+)>\rho^2(E0:\ 2_3^+\rightarrow 2_1^+)$, can be accommodated through adjustments of the Hamiltonian parameters with only minor impact on the energy spectrum. Outside this area the variation is not so abrupt and the values are below one. The area where the ratio takes a value equal to one is surrounding the arc of regularity and, moreover, it forms a quite shallow valley between the critical line and the O(6) symmetry limit. The observed variation for the Xe isotopes is quite small. Note that all Xe isotopes present ratios below one.

Finally, panels c) and f) display $\log\left(\frac{\rho^2(E0:\ 0_2^+\rightarrow 0_1^+)}{\rho^2(E0:\ 2_2^+\rightarrow 2_1^+)}\right)$. In this case, the ratio varies by only about three orders of magnitude. A region with the smallest values of the ratio can be identified, bounded by the arc of regularity and the U(5)-SU(3) leg. The values in this zone are almost constant. The region beyond the arc of regularity displays values
greater than one. It should be noted that, in this situation, the arc aligns much more closely with the Alhassid-Whelan arc of regularity \cite{Alha1991,Alha1991b} than in the other cases shown in Figs.~\ref{fig-ratioCasten_isotopes} and \ref{fig-surfaces_E0}. In the case of Xe isotopes, the ratio goes from values below one at the beginning of the shell to values above one from the middle of the shell till the end.

Therefore, this figure shows that the IBM yields specific ratios tied to particular regions of the Casten triangle that cannot be altered by fine tuning the Hamiltonian parameters and are instead intrinsically determined by the structural character of the nucleus.

\section{Conclusions}
\label{sec-conclu}
In this work, we have investigated how the $\rho^2(E0)$ value can be used to constrain the parameters of the IBM Hamiltonian and have applied this approach to the chain of even-even Xe isotopes. We began by revisiting the relationship between the charge radius and the E0 operators. We then carried out a comprehensive analysis of the even-even Xe nuclei to determine the IBM parameters, using the available experimental data on excitation energies and transition probabilities to fix the parameters of the Hamiltonian and of the E2 transition operator. We also verified that charge radii are properly reproduced. Next, we studied how the matrix elements $|\bra{\psi_f} \hat{n}_d \ket{\psi_i}|^2$ for $0^+$ and $2^+$ states evolve, along with several characteristic $\rho^2(E0)$ ratios. For this purpose, we mapped these quantities across the entire Casten triangle. Finally, for the ratios, we indicated the locations of the Xe isotopes within the Casten triangle on the same plots.

The main outcome of the global analysis over the Casten triangle is that the variation of the $\rho^2(E0)$ values is quite pronounced: there are extended regions where it essentially vanishes, others where it takes finite and relatively uniform values, and still others that show a steep change. This characteristic pattern provides a stringent test of the IBM predictions when nuclei lie in regions with nearly constant values. If the experimental result is not well reproduced there, the agreement cannot easily be improved because the theoretical value is largely insensitive to changes in the Hamiltonian parameters. This situation occurs, for example, for $\rho^2(E0:\ 0_2^+\rightarrow 0_1^+)$ in the vicinity of the symmetry limits.

The Xe isotopes are located in a region close to the QPT line and, in particular, in the second half of the shell they lie relatively near the U(5)–O(6) leg, thus exhibiting an O(5) character. In this area, $\rho^2(E0:0_2^+\rightarrow 0_1^+)$ and $\rho^2(E0:2_3^+\rightarrow 2_1^+)$ display a comparatively pronounced variation, whereas the remaining E0 transitions are almost negligible. As a consequence, access to experimental information will suppose a very stringent test for the IBM results.

In summary, the IBM predicts characteristic $\rho^2(E0)$ values in certain regions of the Casten triangle that are largely insensitive to fine adjustments of the Hamiltonian parameters and are strongly linked to the intrinsic structural character of the nucleus.

\section{Acknowledgments}
This work was partially supported by grant PID2022-136228NB-C21 funded by MCIN/AEI/-10.13039/50110001103 and ``ERDF A way of making Europe''. PMH thanks economical support from grant PREP2022-000423 funded by MCIN/AEI/10.13039/50110001103 and by ``ESF Investing in your future''. Resources supporting this work were provided by the CEAFMC and Universidad de Huelva High Performance Computer (HPC@UHU) funded by ERDF/MI\-NE\-CO project UNHU-15CE-2848.
\bibliography{references-IBM-CM,references-QPT,Refs}

\end{document}